\documentclass[12pt, a4paper]{article}

\usepackage[final]{showkeys}
\usepackage{a4}
\usepackage{amsmath}
\usepackage{amssymb}
\usepackage{epsfig}

\newcommand{\bel}[1]{\begin{equation}\label{#1}}
\newcommand{\bal}[1]{\begin{eqnarray}\label{#1}}

\newcommand{\imag}{\textrm{i}}

\begin{document}

\title{Influence of microscopic transport coefficients
on the formation probabilities for super-heavy elements}
\author{Christian~Rummel\thanks{e-mail:crummel@ph.tum.de} 
\ and \ Helmut~Hofmann \\
 \it{Physik-Department, T39, TUM, D-85747 Garching, Germany}}
\date{July 1st, 2003}
\maketitle

\begin{abstract}
The formation probability is shown to increase by a few orders of
magnitude if microscopic transport coefficients are used rather
than those of the common macroscopic pictures. Quantum effects in
collective dynamics are taken into account through the fluctuating
force, as exhibited in diffusion coefficients for a Gaussian
process. In the range of temperatures considered here, they turn
out to be of lesser importance.

\vspace{5mm}
\end{abstract}

\section{Introduction}
\label{intro}

The description of fusion of heavy ions is commonly performed by
studying  separately the entrance phase and the formation of the
compound nucleus
\cite{ann.che.naa.pev.vov:prc:95}-\cite{ary.ohm:pan:03}.
Typically, for the latter process the system has to overcome a
barrier and in this sense resembles nuclear fission. Both
processes may be described by transport models involving
Fokker-Planck or Langevin equations. One essential difference of
both cases is found in different initial and boundary conditions.
Whereas fission can be pictured as the decay of a meta-stable
state the formation process starts out from some intermediate
distribution in collective phase space. Often the latter is simply
assumed to be one which moves towards the inner barrier, a picture
which we will adopt in this paper. Our main aim will be to examine
the importance of using microscopic transport coefficients which
in one way or other are governed by quantum effects. To this end
we are going to compare such calculations with results obtained
within the commonly assumed macroscopic picture. For thermal
fission the influence of microscopic transport coefficient has
been shown to modify the decay rate considerably
\cite{hoh.ivf:prl:99,hoh.ivf.ruc.yas:prc:01}.

\section{The formation probability}
\label{formation}

\begin{figure}[h] \begin{center}
\epsfig{file=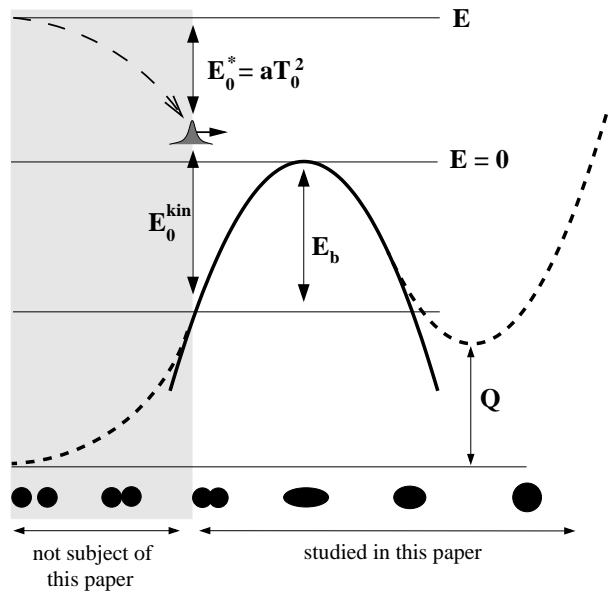, width=80mm}
\caption{\label{fig-scenario} Sketch of the scenario of this
paper.}
\end{center} \end{figure}

We assume the formation of the compound nucleus to be described as
a Gaussian Markov process over an inner barrier, see
Fig.~\ref{fig-scenario}.
In order to allow for a largely
analytic approach the barrier will be supposed to be given by an
inverted oscillator. Restricting ourselves to Gaussian
distributions in phase space the probability to overcome this
barrier is simply given by \cite{hoh.sip:plb:75}
\bel{formprob-tdep} \Pi(t) = \frac{1}{2} \left[1 + \textrm{erf}
\left(\frac{q(t)}{\sqrt{2\Sigma_{qq}(t)}}\right)\right] =
\frac{1}{2} \,\textrm{erf} \left(
-\frac{q(t)}{\sqrt{2\Sigma_{qq}(t)}} \right)
\end{equation}
The $q(t) = \langle q \rangle_{t}$ specifies the center of the
Gaussian, which moves from left to right and which within this
approximation represents the average trajectory. The
$\Sigma_{qq}(t) = \langle (q - \langle q \rangle_{t})^{2}
\rangle_{t}$ defines the fluctuation of the coordinate. Analogous
definitions are used for the kinetic momentum $P$, both for the
average trajectory $P(t)$ as well as for the fluctuations
$\Sigma_{pp}(t)$ and $\Sigma_{qp}(t)$. Analytical expressions for
$q(t)$ and $\Sigma_{qq}(t)$, which may be derived from a
Fokker-Planck equation \cite{hoh:pr:97}, are given in appendix
\ref{timeevol}. Fortunately, when time progresses past a certain
"relaxation time" ($t \gg 1/|z^{-}|$) the probability
(\ref{formprob-tdep}) approaches a finite stationary value:
\bel{formprob} \Pi_{\infty} = \lim_{t\to\infty} \Pi(t) =
\frac{1}{2} \,\textrm{erf} \left( -\frac{q_{\infty}}
{\sqrt{2\Sigma_{qq}^{\infty}}} \right) \,.
\end{equation}
The time dependence of both the coordinate as well as of the
fluctuation is determined by the frequencies
(see appendix \ref{timeevol})
\bel{zpm} z^{\pm} = \frac{1}{2\tau_\textrm{kin}} \left( \pm
\sqrt{1 + \frac{1}{\eta_b^{2}}} - 1\right)\,.
\end{equation}
The  $\tau_\textrm{kin}$ measures the relaxation of the momentum
to the thermal Maxwell distribution and the (dimensionless)
parameter $\eta_b$ indicates whether the process is under-damped
($\eta_b < 1$) or over-damped ($\eta_b > 1$). In terms of the
transport coefficients for average motion, inertia $M$, friction
$\gamma$ and stiffness $C$, which here is negative $C = -|C|$,
these quantities are given by
\bel{tkineta}\tau_\textrm{kin} = \frac{M}{\gamma}   \qquad
\text{and} \qquad \eta_b = \frac{\gamma}{2\sqrt{|C|M}} =
\frac{1}{2} \sqrt{\frac{\tau_\textrm{coll}}{\tau_\textrm{kin}}}\,.
\end{equation}
For strong damping $\eta_b \gg 1$,
viz $\tau_\textrm{coll} \gg \tau_\textrm{kin}$, we have
\bel{zplus-OVDA}
z^{+} \longrightarrow
\frac{1}{\tau_\textrm{coll}} \equiv \frac{|C|}{\gamma}
\qquad \text{and} \qquad
z^{-} \longrightarrow -\frac{1}{\tau_\textrm{kin}} \,.
\end{equation}
In this limit only processes are relevant (see (\ref{sol-q}) and
(\ref{sol-fluct})) which take place on the time scale
$\tau_\textrm{coll}$.
Later on we will also need the frequency $\varpi_{b}$ at the barrier,
\bel{freqbar}
\varpi_{b} = \sqrt{\frac{|C|}{M}} =
\frac{1}{\sqrt{\tau_\textrm{coll} \tau_\textrm{kin}}} \,.
\end{equation}

The dynamics of fluctuations is not only controlled by the
transport coefficients of average motion but by diffusion
coefficients as well. In the classical limit (for collective
motion) there is only the one given by the famous Einstein
relation $D_{pp}= \gamma T$. In the quantum case and for larger
damping there may be another one, $D_{qp}$. Both are to be
calculated from the quantal fluctuation-dissipation
theorem\footnote{It is for this reason that there are no more than
2 diffusion coefficients}, if the latter only is generalized to
the barrier problem by a suitable analytic continuation
\cite{hoh.kid:ijmpe:98, hoh:pr:97}.  We will address this question
in more detail later in sect. \ref{statfluct}.

Finally, we like to note that, different to the original papers,
we want to use dimensionless coordinates and the conjugate
momenta of dimension $\hbar$. 
This means to apply the following canonical replacement
\bel{dimless-QP}
q \rightarrow \sqrt{\frac{|C|}{\hbar\varpi_b}} \,q \ , \qquad
P \rightarrow \sqrt{\frac{\hbar}{M\varpi_b}} \,P \,,
\end{equation}
which for the second moments implies
\bel{dimless-fluct}
\Sigma_{qq} \rightarrow \frac{|C|}{\hbar\varpi_{b}} \,\Sigma_{qq} \ ,
\quad
\Sigma_{pp} \rightarrow \frac{\hbar}{M\varpi_b} \,\Sigma_{pp} \ ,
\quad
\Sigma_{qp} \rightarrow \Sigma_{qp} \,.
\end{equation}

\section[Transport coefficients]
{Transport coefficients of average motion and diffusion}
\label{transportCOE}

To distinguish between the coefficients of average motion, $M$,
$\gamma$ and $C$, and the diffusion coefficients $D_{\mu\nu}$, the
notion transport coefficients will henceforth be reserved for the
former. Later on we are going to compare results obtained for
microscopic and macroscopic pictures of the dynamics of nucleons
inside the nucleus. In the first case this refers to {\em quantal}
motion of nucleons in a mean field. Within the Gaussian approach
quantum effects in collective motion only show up in the diffusion
coefficients and, hence, only in the dynamics of the fluctuations.
With respect to this property the notion macroscopic limit stands
synonymous for the high temperature limit for which the Einstein
relation applies.

\subsection{Transport coefficients of average motion}
\label{tran-avmo}

\begin{figure}[h]
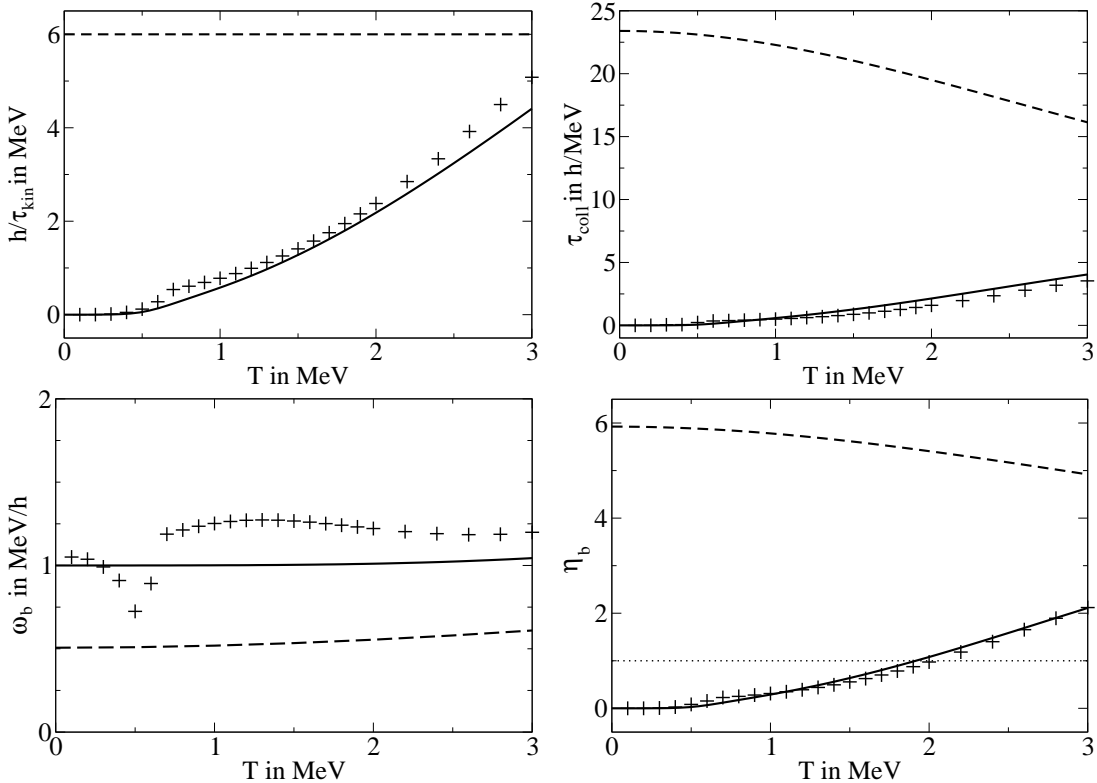
 \begin{center}
\epsfig{file=figure2a.eps, width=0.49\linewidth} \hfill
\epsfig{file=figure2b.eps, width=0.49\linewidth} \\
\epsfig{file=figure2c.eps, width=0.49\linewidth} \hfill
\epsfig{file=figure2d.eps, width=0.49\linewidth}
\caption{\label{fig-transcoeff} Temperature dependence of relevant
ratios of transport coefficients. Critical damping $\eta_b = 1$ is
marked by a dotted line in the bottom right panel. Crosses refer
to the microscopic computations of \cite{hoh.ivf.ruc.yas:prc:01}
(with pairing included). The solid line represents the fits
(\ref{taukin}) and (\ref{taucoll}) to these data. Macroscopic
transport coefficients are given by the dashed line
\cite{ivf.hoh.pav.yas:prc:97}.}
\end{center} \end{figure}

\paragraph{Microscopic model:}
Concerning the microscopic transport coefficients we use the
suggestion made in \cite{hoh.ivf.ruc.yas:prc:01}
\begin{eqnarray}
\tau_\textrm{kin} & =  & \frac{1 + T^{2}/40}{0.6 T^{2}}
\ \frac{\hbar}{~\textrm{MeV}} \label{taukin} \\
\tau_\textrm{coll} & = & \frac{0.6 T^{2}} {1 +
\pi^{2}T^{2}/c_\textrm{macro}^{2}}  \ \frac{\hbar}{~\textrm{MeV}}
\label{taucoll}
\end{eqnarray}
(with temperature being measured in MeV). 
These simple expressions have been developed to resemble the
$T$-dependence of the truly microscopic values of
$\tau_\textrm{kin}(Q,T)$ and $\tau_\textrm{coll}(Q,T)$ 
obtained by application of linear response theory. 
Using (\ref{taukin}) and (\ref{taucoll}) we calculate the
effective damping strength $\eta_b(T)$ from (\ref{tkineta}) and
the frequency $\varpi_{b}(T)$ from (\ref{freqbar}). As may be seen
from Fig.~\ref{fig-transcoeff} the latter may fairly well be
approximated by $\varpi_{b}(T)\approx 1 ~\textrm{MeV}/\hbar$. In
(\ref{taucoll}) the $c_\textrm{macro}$ is introduced as a
parameter to fix the high temperature limit of friction. In more
recent times this is claimed to be given not by the wall formula
$\gamma_\textrm{wall}$ but by about one half of this value, see
e.g. \cite{hoh.ivf.ruc.yas:prc:01} and further references given
there; we adopt this suggestion. It may be mentioned in passing
that the estimates given in (\ref{taukin}) and (\ref{taucoll})
could be extended to include the influence of pairing
\cite{hoh.ivf.ruc.yas:prc:01}. The latter diminishes friction at
lower temperatures $T \lesssim 0.5~\textrm{MeV}$ 
but its influence is negligible in the range of
temperatures used in the present application.

\paragraph{Macroscopic model:}
In \cite{shc.kog.aby:prc:02} the friction coefficient is taken
from the wall-and-window formula. Like in
\cite{hoh.ivf.ruc.yas:prc:01} we will use
$\gamma = \gamma_\textrm{wall}/2$ here. The macroscopic value of the
frequency $\varpi_{b}(T)$ can be calculated from the liquid drop
values of stiffness $C_\textrm{LDM}(T)$ and inertia
$M_\textrm{LDM}$; we followed the procedure described in
\cite{ivf.hoh.pav.yas:prc:97}. The ratios $\tau_\textrm{kin}(T)$
and $\eta_b(T)$ are then fixed by (\ref{tkineta}).

In Fig.~\ref{fig-transcoeff} we show a comparison of the
temperature dependence of the microscopic and macroscopic
transport coefficients used in this paper. It is most important to
realize that in the microscopic case collective motion changes
from under-damped to over-damped at a temperature $T \approx 2
~\textrm{MeV}$, whereas in the macroscopic limit collective motion
is over-damped ($\eta_{b} > 1$) even at very small temperatures.
Please have in mind that with increasing $\eta_b$ the inertia $M$
becomes less and less important, to drop out entirely for the
truly over-damped case. For this reason the estimates given above
should be understood to represent those quantities  which involve
$M$ only for smaller temperatures, up to $T \approx 2 \ldots 3
~\textrm{MeV}$ at best.

\subsection{Diffusion coefficients}
\label{statfluct}

As mentioned already before, the diffusion coefficients are to be
calculated from the fluctuation dissipation theorem. 
The latter specifies equilibrium fluctuations
$\Sigma_{\mu\nu}^{eq}$ to which the dynamical ones turn to at
large times (see appendix \ref{diffcoeff-FDT} or \cite{hoh:pr:97}). 
To evaluate the $\Sigma_{\mu\nu}^{eq,b}$ for a
barrier, as represented by an inverted oscillator, requires a
suitable analytic continuation involving the change of the
stiffness from $|C|$ to $C = -|C|$; details of the procedure may
be found in \cite{hoh.kid:ijmpe:98,hoh:pr:97}. Accounting for our
presently used transformation (\ref{dimless-fluct}) the relations 
to the diffusion coefficients become
\bel{fluct-diff} D_{qq} = \Sigma_{qp}^{eq,b} = 0 \ , \quad 
D_{qp} = \hbar\varpi_b \,(|\Sigma_{qq}^{eq,b}| - 
\Sigma_{pp}^{eq,b}/\hbar^{2}) \ , \quad 
D_{pp} = \tau_\textrm{kin}^{-1} \,\Sigma_{pp}^{eq,b}
\end{equation}
Please note that the $\Sigma_{qq}^{eq,b}$ of the inverted oscillator
is negative.

The application of (\ref{fluct-diff}) becomes particularly simple
in the {\em high temperature limit}. The classic equipartition theorem
of Statistical Mechanics implies that  one has
\bel{Seqclass}
|\Sigma_{qq}^{eq,b}| = \Sigma_{pp}^{eq,b}/\hbar^{2} =
\frac{T}{\hbar\varpi_{b}} \,,
\qquad
\Sigma_{qp}^{eq,b} = 0
\qquad \text{for} \qquad
T \gg \hbar\varpi_{b} \,.
\end{equation}
This is easily recognized considering (\ref{dimless-fluct}) again.
As a consequence, the diffusion coefficients turn to the Einstein
relations, which in our case read
\bel{Einstein}
D_{pp}^{h.T.} =
\frac{1}{\tau_\textrm{kin}} \,\frac{\hbar T}{\varpi_{b}}
= 2\hbar \eta_b T \,, \qquad
D_{qq}^{h.T.} = D_{qp}^{h.T.} = 0 \,.
\end{equation}
For very small damping formulas (\ref{Seqclass}) and
(\ref{Einstein}) keep their form even in the quantum case if
only $T$ is replaced by the effective temperature
\bel{Teff} T_{b}^{*} = \frac{\hbar\varpi_{b}}{2} \,\cot \left(
\frac{\hbar\varpi_{b}}{2T} \right)\,,
\end{equation} see
\cite{pok.hoh:jp:81} or \cite{hoh:pr:97}. As this requires $\eta_b
\ll 1$ this limit sometimes is called the {\em zero damping
limit}.
It is seen that friction neither appears in the classical or high
temperature limit  (\ref{Seqclass}) nor, by definition, in the
limit (\ref{Teff}) of zero damping.

\begin{figure}[h]
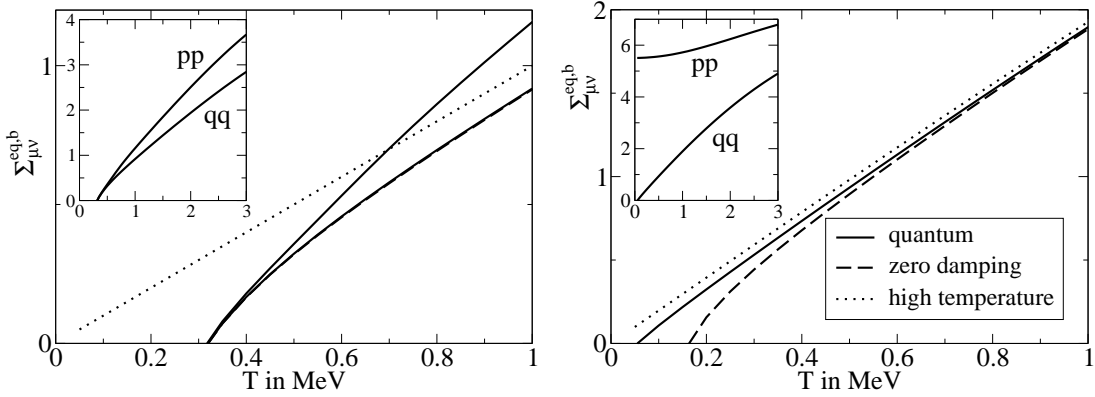
 \begin{center}
\epsfig{file=figure3a.eps, width=0.49\linewidth} \hfill
\epsfig{file=figure3b.eps, width=0.49\linewidth}
\caption{\label{fig-eqfluct} $T$-Dependence of the equilibrium
fluctuations of coordinate and momentum calculated in various limits
for microscopic (left) and macroscopic transport coefficients (right).
The inserts show the behavior at higher temperatures.}
\end{center} \end{figure}

At lower temperature this is no longer true for the general
situation where the size of damping plays an essential role on the
temperature dependence of the $\Sigma_{\mu\nu}^{eq,b}$, and hence
on the diffusion coefficients. This has already been demonstrated
in \cite{hoh.kid:ijmpe:98,hoh:pr:97}. In these calculations the
transport coefficients of average motion simply appeared as given,
constant parameters. As discussed above, they themselves may vary
sensitively with $T$. For this reason the calculations of
\cite{hoh.kid:ijmpe:98} have been repeated using now the specific
transport coefficients for average motion described above. The
results for $|\Sigma_{qq}^{eq,b}|$ and $\Sigma_{pp}^{eq,b}$ are
shown in Fig.~\ref{fig-eqfluct}. It must be said that the
calculation of the momentum fluctuations requires regularization
of a frequency integral, which comes in through the fluctuation
dissipation theorem (\ref{FDT}). Also with respect to this problem 
the same procedure has been applied as in \cite{hoh.kid:ijmpe:98} 
using a Drude regularization with $\omega_\textrm{D} \approx 
10 \varpi$ as the relevant cut-off parameter. As we have
$\varpi_{b}(T) \approx 1 ~\textrm{MeV}/\hbar$ (see bottom left
panel of Fig.~\ref{fig-transcoeff}) we took the cut-off parameter
$\omega_\textrm{D} = 10 ~\textrm{MeV}/\hbar$ after convincing
ourselves that the results are not extremely sensitive to this
choice. The variation with $\omega_\textrm{D}$ is more pronounced
in the case of macroscopic transport coefficients where for
$\omega_\textrm{D} = 5 \ldots 20 ~\textrm{MeV}/\hbar$ the momentum
fluctuations $\Sigma_{pp}^{eq,b}$ differ from our choice by an
additional factor of $0.8 \ldots 1.3$ at $T = 1 ~\textrm{MeV}$. In
the microscopic case the corresponding factor is only $0.8 \ldots
1.1$. Paying attention to the fact that the effective damping is
quite different in both cases, this is in good agreement with the
findings of \cite{hoh.kid:ijmpe:98}.

In Fig.~\ref{fig-eqfluct} the $T$-dependence of
$\Sigma_{\mu\nu}^{eq,b}$ is exhibited both for the microscopic
transport coefficients (left panel) as well as for those in the
macroscopic limit (right panel), as described in
Sect.~\ref{tran-avmo}. In each case the fully quantum result is
compared with the limits of high temperature (classical) and zero
damping.
It is seen that at very small temperatures the microscopic results
are very well represented by the limit of weak damping.
For $\Sigma_{pp}^{eq,b}$ the deviations grow
faster, a feature which might by related to the fact that the
regularization procedure was taken over from
\cite{hoh.kid:ijmpe:98} without modification. This problem has not
been studied any further, simply because the effect of the
$\Sigma_{pp}^{eq,b}$ on the transmission probability is weakened
by other effects, in particular at larger damping, as will be
explained shortly. In any case, the effect seen at small $T$
simply results from the fact that the microscopic damping strength
$\eta_b$ becomes {\em very small} (see lower right part of
Fig.~\ref{fig-transcoeff}). Let us note that at large temperatures
the microscopic and macroscopic diffusion coefficients approach
each other, if normalized to $\hbar\varpi_{b}(T)$,
simply because this is so for the transport coefficients of
average motion discussed in Sect.~\ref{tran-avmo}.

\section{Numerical results}
\label{numerics}

In the last two subsections it was shown that a proper microscopic
treatment of the quantum heat bath has two major effects: (i) the
effective damping strength $\eta_{b}$ becomes much smaller than that of
macroscopic models and (ii) different to the Einstein relation 
a second diffusion coefficient $D_{qp}$ shows up. After
specifying the initial conditions for our numerical calculations 
we will examine these effects separately. The effect of quantum
diffusion on the formation probabilities (\ref{formprob-tdep}) 
and (\ref{formprob}) is contrasted with the limits of high temperature 
(vanishing $D_{qp}$) and zero damping. This will be done both for 
microscopic as well as macroscopic transport coefficients of average
motion.

\subsection{Initial conditions}
\label{inicond}

In terms of the barrier height and the initial kinetic energy, 
expressed by
\bel{def-EbK0}
E_{b} = -V_{0} = \frac{\hbar\varpi_{b}}{2} \,q_{0}^{2}
\qquad \textrm{and} \qquad
E_{0}^\textrm{kin} = \frac{\varpi_{b}}{2\hbar} \,P_{0}^{2}
\end{equation}
through the initial values of the average
coordinate $q_{0}$ and the momentum $P_{0}$,
the argument of the error function in (\ref{formprob}) turns into
\bel{errarg}
\frac{q_\infty}{\sqrt{2 \Sigma_{qq}^{\infty}}} =
\frac{\sqrt{E_{0}^\textrm{kin}/\hbar\varpi_{b}} + \sqrt{E_{b}/\hbar\varpi_{b}}
\,(z^{-}/\varpi_{b})} {\sqrt{(\Sigma_{pp}(t_{0}) -
\Sigma_{pp}^{eq,b})/\hbar^{2} + (z^{-}/\varpi_{b})^{2} \,(\Sigma_{qq}(t_{0})
- \Sigma_{qq}^{eq,b}) - 2 (z^{-}/\varpi_{b})
\,\Sigma_{qp}(t_{0})/\hbar}} \,
\end{equation}
see appendix \ref{timeevol}.
A very similar result has been obtained by application of Langevin
dynamics to the inverted oscillator \cite{aby.bod.gib.wat:pre:00}.

\begin{figure}[hbt] \begin{center}
\epsfig{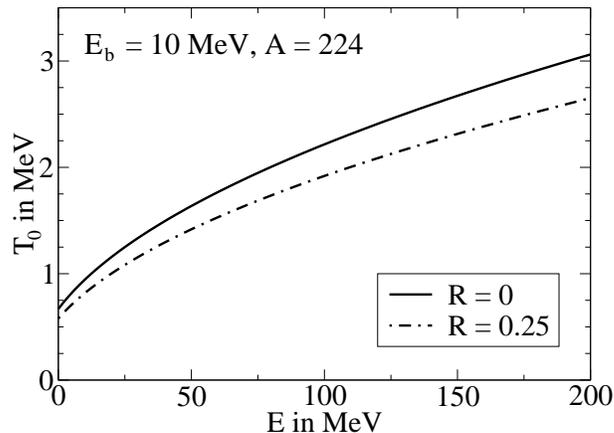} \caption{\label{fig-T0}
Dependence of the initial temperature on the incident energy.}
\end{center} \end{figure}

For the barrier height specified in (\ref{def-EbK0}) we choose
$E_{b} = 10 ~\textrm{MeV}$. This is in fair agreement with the
typical heights of the inner barriers used in
\cite{dev.hos:prc:00}. The initial kinetic energy $E_{0}^\textrm{kin}$ and the
temperature $T_{0}$ depend on the strength of friction in the
approach phase. As in the present paper this problem is not
addressed in detail we will simply assume the incident energy $E$
over the barrier top to be strongly dissipated such that a
scenario holds as outlined in Fig.~\ref{fig-scenario}. As can be
seen there, the sum of the initial energy $E$ and the barrier
height $E_{b}$ are split into kinetic energy and thermal
excitation according to
\bel{branchE}
E_{0}^\textrm{kin} = R (E + E_{b})
\qquad \textrm{and} \qquad
E_{0}^{*} = a T_{0}^{2} = (1 - R) (E + E_{b}),
\end{equation}
with the ratio $0 \le R \le 1$ to be fixed later. As an estimate
for the level density parameter we use the value $a \approx A / 10
~\textrm{MeV}$ where $A$ is the nucleon number of the compound
nucleus. In order to keep the computation simple, we assume that
temperature stays constant at its initial value $T = T_{0}$
throughout the whole formation process. In Fig.~\ref{fig-T0} we
show the dependence of $T_{0}$ on the energy $E$ for a system of
nucleon number $A = 224$ and two different assumptions for $R$. 
In \cite{ary.wat.ohm.aby:prc:99}
temperatures in the range $T_{0} = 0.68 \ldots 1.24 ~\textrm{MeV}$
have been studied for  a heavier system with $A = 298$ nucleons.
On the basis of the surface friction model it has been argued 
that almost no kinetic energy is left after the approach phase, 
viz at the starting point $q_{0}$ (see for instance Fig.~3 of
\cite{aby:epja:02} and Fig.~1b of \cite{shc.kog.aby:prc:02}). 
In contrast the authors of \cite{kog.shc.aby:jnrs:02,ary.ohm:pan:03} 
have suggested damping in the approach phase to be weaker  
than that of surface friction. In this work we do not want to
elaborate on the approach phase but simply incorporate this 
uncertainty by comparing the case 
of vanishing initial kinetic energy $E_{0}^\textrm{kin} = 0$, for
which the parameter of (\ref{branchE}) becomes $R = 0$, with
the one of a weaker friction force at large elongations, 
chosen such that $R=0.25$.

To specify the Gaussian of the initial
phase space distribution we need the widths in $q$ and $P$. They
shall be parameterized in terms of the $\Sigma_{pp}^{eq,b}$ by
introducing two positive numbers $w$ and $W$,
\bel{S-ini} \Sigma_{pp}(t_{0}) = w \,\Sigma_{pp}^{eq,b} \,, \quad
\Sigma_{qq}(t_{0}) = W \,\frac{\hbar^{2}}{4 w \Sigma_{pp}^{eq,b}} \,,
\quad \Sigma_{qp}(t_{0}) = 0\,.
\end{equation}
The $w$ fully determines the momentum fluctuation and the $W$ the
one in the coordinate provided the former is given.  In case of a
strong friction force in the approach phase the $w$ will be close
to unity, as then the momentum distribution will be close to
Maxwell distribution \cite{hoh:pr:97}. The $W$ can be said to
define a kind of measure for the overall width of the phase space
distribution. Indeed, requiring $W \ge 1$ warrants the uncertainty
principle
\bel{uncert}
\langle (q - \langle q \rangle_{t_{0}})^{2} \rangle_{t_{0}}
\langle (P - \langle P \rangle_{t_{0}})^{2} \rangle_{t_{0}} =
\Sigma_{qq}(t_{0}) \,\Sigma_{pp}(t_{0}) \ge \hbar^{2}/4
\end{equation}
to be fulfilled initially. In the limiting case $W = 1$ the system
starts out of the most narrow distribution still compatible with
quantum mechanics. As we want to examine the importance of genuine
quantum effects vanishing initial fluctuations
$\Sigma_{\mu\nu}(t_{0}) \equiv 0$ are prohibited for our studies.
One must have in mind, however, that in reality finite fluctuations
will have been built up in the approach phase, which may not always be
represented by simple Gaussians. In our numerical calculations,
the $\Sigma_{pp}^{eq,b}$ which enters the initial conditions
(\ref{S-ini}) is always taken to be the quantum value (see Sect.
\ref{statfluct}), even for studies of the time evolution within
high temperature limit.

From the results displayed in Fig.~\ref{fig-eqfluct} it can be
seen that the $\Sigma_{\mu\nu}^{eq,b}$ of the zero damping and
high temperature limits may differ considerable from those of the
full, quantum results. This difference is particularly large for
$\Sigma_{pp}^{eq,b}$ when macroscopic transport coefficients
of average motion are used. One might be inclined to believe
that at small temperatures this might have a large impact on
the formation probabilities $\Pi(t)$ and $\Pi_\infty$.
However, one must be aware that not only the
$\Sigma_{\mu\nu}^{eq,b}$ contribute to the $\Sigma_{qq}^{\infty}$
of the denominator of (\ref{errarg}) but the initial fluctuations
$\Sigma_{\mu\nu}(t_{0})$ as well. To reveal these effects
\begin{figure}[h]
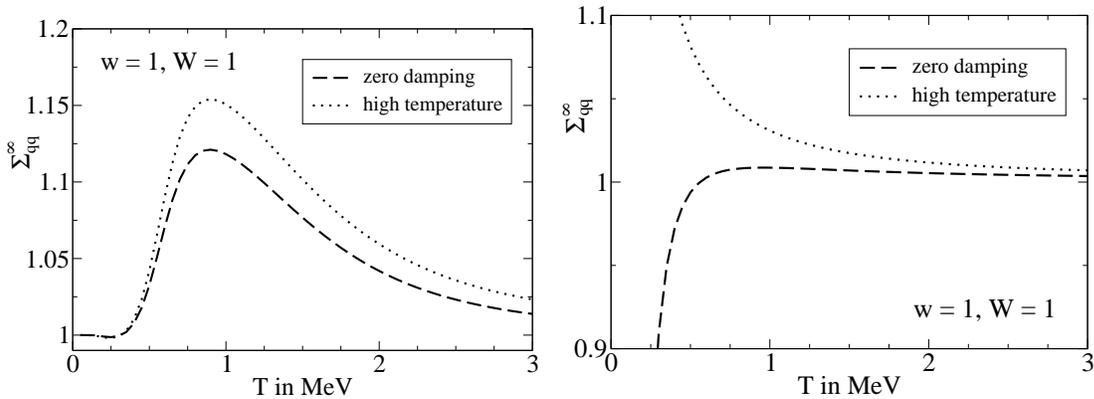
 \begin{center}
\epsfig{file=figure5a.eps, width=0.49\linewidth} \hfill
\epsfig{file=figure5b.eps, width=0.49\linewidth}
\caption{\label{fig-alpha2} Temperature dependence of the
$\Sigma_{qq}^{\infty}$ for the zero damping and high temperature
limit, both divided by the general result: left panel for
microscopic input, right panel for macroscopic input.}
\end{center} \end{figure}
we address in Fig.~\ref{fig-alpha2} the $T$-dependence of the
$\Sigma_{qq}^{\infty}$. Shown there is the ratio of its values in
either the high temperature limit or the zero damping limit to
that for the full calculation, for the choice $w = 1$ and $W = 1$
--- for other cases this effect is even smaller.
It is seen that the values $\Sigma_{qq}^{\infty}$ obtained in
these calculations are quite close to each other. Note, please,
that in our studies only temperatures $T \gtrsim 0.6
~\textrm{MeV}$ are accessible (see Fig.~\ref{fig-T0}).  As only
the square root of the $\Sigma_{qq}^{\infty}$ appears in
(\ref{formprob}) one may already deduce here that the three
calculations will lead to results which do not differ too much.

\subsection{Formation probabilities for microscopic input}
\label{microresults}

\begin{figure}[hbt] \begin{center}
\epsfig{file=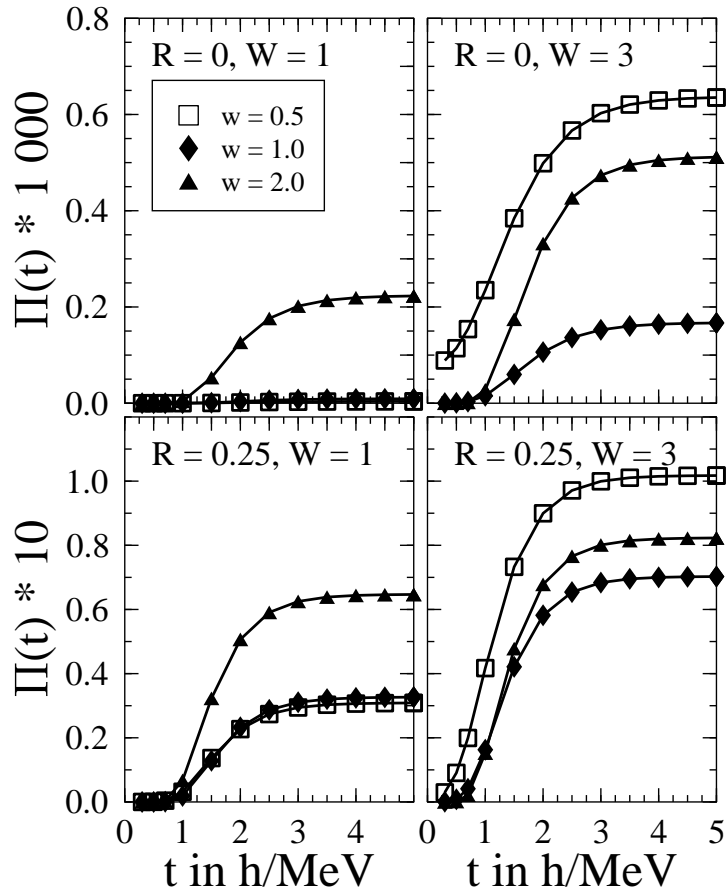, width=0.7\linewidth}
\caption{\label{fig-micro-Pioft} Time evolution of the formation
probability $\Pi(t)$ for microscopic transport coefficients and 
quantum diffusion coefficients, for different choices of the parameters 
specified in the inserts; 
the incident energy is $E = 10~\textrm{MeV}$. 
Note that the scaling in the upper and lower panels differs by two 
orders of magnitude. For more details see text.}
\end{center} \end{figure}

We plot the time evolution of the
formation probability using the microscopic set of
transport coefficients and the fully quantum calculation of the
diffusion coefficients in Fig.~\ref{fig-micro-Pioft}. 
Results are shown for an incident energy
above the barrier of $E = 10~\textrm{MeV}$ and for vanishing
($R = 0$, $E_{0}^\textrm{kin} = 0$) and finite initial kinetic energy
($R = 0.25$, $E_{0}^\textrm{kin} = 5~\textrm{MeV}$).
For the parameter $w$, which according to
(\ref{S-ini}) specifies the initial width in the momentum, three
different values are used: $w = 1/2$ (more narrow than thermal
equilibrium), $w = 1$ (thermal equilibrium) and $w = 2$ (wider
than thermal equilibrium). The initial distribution in the 
coordinate is specified by the two values $W = 1$ and $W = 3$. 
In most cases the initial space distribution is almost totally
located on the left of the barrier. For the choice $w = 1/2$ and
$W = 3$ a small part of the distribution is on its right 
already at the beginning. In this case the formation probability 
$\Pi(t)$ does not start from exactly zero. 

In Fig.~\ref{fig-micro-Pioft} we show examples corresponding to small
effective damping strengths: $\eta_b = 0.26$ and $\eta_b = 0.18$ at 
$T = 0.94~\textrm{MeV}$ ($R = 0$) and $T = 0.82~\textrm{MeV}$ 
($R = 0.25$), respectively. As this corresponds
to  under-damped motion the initial kinetic energy plays an
important role on the formation probability. Its stationary value
$\Pi_\infty$  increases by two orders of magnitude in going from
$R = 0$ to $R = 0.25$. 
Repeating the same calculations at larger energy $E = 50~\textrm{MeV}$ 
reveals additional effects. Enlarging $R$ from $0$ to $0.25$, 
which is to say for increasing initial kinetic energy and slightly 
decreasing temperature 
($E_{0}^\textrm{kin} = 0$, $T_{0} = 1.6~\textrm{MeV}$ 
in the case $R = 0$ and $E_{0}^\textrm{kin} = 15~\textrm{MeV}$, 
$T_{0} = 1.4~\textrm{MeV}$ in the case $R = 0.25$), 
the asymptotic value $\Pi_\infty$ 
of $\Pi(t)$ is reached quicker, independently of $w$ and $W$. 
The influence of the parameters $W$ and $w$ on the value of $\Pi_\infty$ 
is less important than the one of $R$: Varying either $w$ or $W$
mostly implies a change by less than one order of magnitude.

\begin{figure}[hbt]
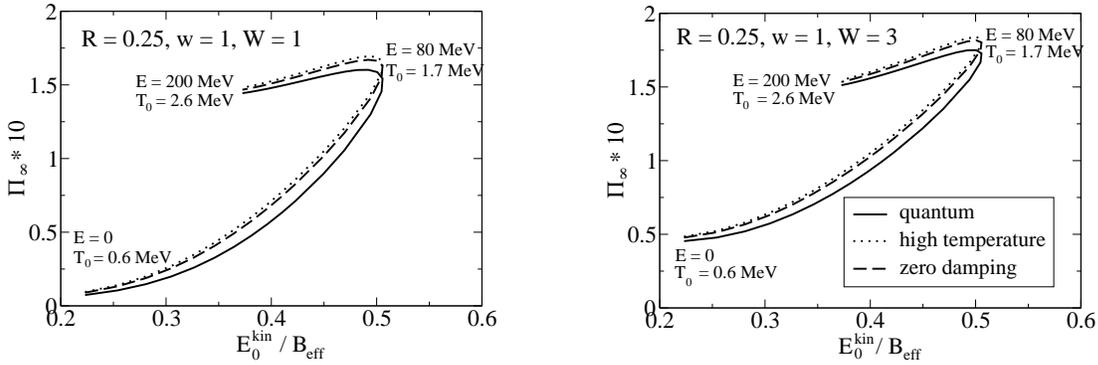
 \begin{center}
\epsfig{file=figure7a.eps, width=0.45\linewidth} \hfill
\epsfig{file=figure7b.eps, width=0.45\linewidth}
\caption{\label{fig-limPiofE-micro} Formation probability
$\Pi_\infty$ using microscopic transport coefficients as a
function of the ratio of the initial kinetic energy
(\ref{branchE}) to the effective barrier height (\ref{barreff}).
For $\Sigma_{\mu\nu}^{eq,b}$ we compare the results applying a
quantum calculation (comp. \cite{hoh.kid:ijmpe:98}) and the limits
of high temperature (\ref{Seqclass}) and zero damping
(\ref{Seqclass}) with (\ref{Teff}). Note that the incident energy
changes along the curve: $E = 0 \ldots 200~\textrm{MeV}$. 
The value $E = 10~\textrm{MeV}$ corresponds to
$E_{0}^\textrm{kin}/B_\textrm{eff} = 0.35$;  
for details see text.}
\end{center} \end{figure}

Next we discuss the energy dependence of the asymptotic value
$\Pi_\infty$. This is done best by introducing an ``effective barrier
height'' by the following argument: 
Within the Gaussian approximation the transmission across the
barrier is largely governed by the average trajectory. As can be
seen from (\ref{formprob}) $\Pi_{\infty}$ becomes equal to $1/2$
when the trajectory just hits the top of the barrier at $q_\infty
= 0$. By inspecting the numerator of (\ref{errarg}), and recalling
from eq.(\ref{zpm})  the value of $z^-$,  one easily recognizes
that the trajectories may be classified by their initial kinetic
energy $E_{0}^\textrm{kin}$ in comparison with the "effective
barrier height" (see also
\cite{hoh.sip:plb:75,aby.bod.gib.wat:pre:00,aby:epja:02})
\bel{barreff}
B_\textrm{eff} =
\left( \sqrt{1 + \eta_b^2} + \eta_b \right)^2 \,E_{b} \ge E_{b} \,.
\end{equation}
Trajectories with $E_{0}^\textrm{kin} < B_\textrm{eff}$ are simply
reflected implying the $\Pi_{\infty}$ to be smaller than $1/2$;
 only those trajectories with $E_{0}^\textrm{kin} > B_\textrm{eff}$
are able to cross the barrier and the $\Pi_\infty$ exceeds $1/2$.
Fig.~\ref{fig-limPiofE-micro} shows the functional dependence of
$\Pi_\infty$ on the initial kinetic energy scaled to the effective
barrier height, for $R = 0.25$, $w = 1$ and $W = 1$ (left) and $W
= 3$ (right). 
For several reasons the incident energy is restricted to the interval 
$E = 0 \ldots 200~\textrm{MeV}$, with a corresponding temperature of 
$T_{0} = 0.6 \ldots 2.6~\textrm{MeV}$. First of all, at larger
temperatures the survival probability becomes too small. Secondly, as
mentioned already before, there the motion becomes strongly
over-damped, such that formulas which involve the inertia become less
and less trustworthy.
At small incident energies $E$ (small $T_{0}$), for $\Pi_\infty$ a 
choice of $W = 3$ implies a factor of the order $5$ over the case 
of the most narrow initial distribution which still is 
compatible with quantum mechanics ($W = 1$). This increase of
$\Pi_\infty$ with $W$ rapidly diminishes with growing $E$. In
addition to the full quantum calculation of the diffusion
coefficients we show in Fig.~\ref{fig-limPiofE-micro} the results
of the limits of zero damping and high temperature. Both
approximations are of about the same quality. For the chosen example
quantum effects are seen to be small and become largest at
intermediate temperatures $T = 0.7 \ldots 2~\textrm{MeV}$ ($E = 5
\ldots 120~\textrm{MeV}$). Both observations confirm the
conjecture made already in connection with Fig.~\ref{fig-alpha2}.

As a remarkable feature we observe in
Fig.~\ref{fig-limPiofE-micro} a back-bending of the curves
representing $\Pi_\infty$ at $E_{0}^\textrm{kin}/B_\textrm{eff}
\approx 0.5$. It can be explained as follows:
The initial kinetic energy depends on $E$ linearly 
(see (\ref{branchE})), while the $E$-dependence of the effective 
barrier height (\ref{barreff}) is more complicated and comes in via 
$\eta_{b}(T_{0}(E))$. As a function of $E$ for $R = 0.25$ the ratio
$E_{0}^\textrm{kin}/B_\textrm{eff}$ develops a maximum and a minimum
(the latter corresponding to $T_{0} \approx 4~\textrm{MeV}$; which is
beyond the region considered here). At energies far beyond 
this range $\Pi_{\infty}$ would approach $1$. The final formation
probability also exhibits a shallow maximum near the turning
point. This behavior might have some influence on the reasonable
choice of incident energies in an experimental setup for heavy ion
collisions, because the survival probability decreases with
increasing excitation energy $E^{*}$ and therefore smaller $E$ is
more favorable for survival.

\subsection{Formation probabilities for macroscopic coefficients of
average motion}
\label{macroresults}

\begin{figure}[hbt] \begin{center}
\epsfig{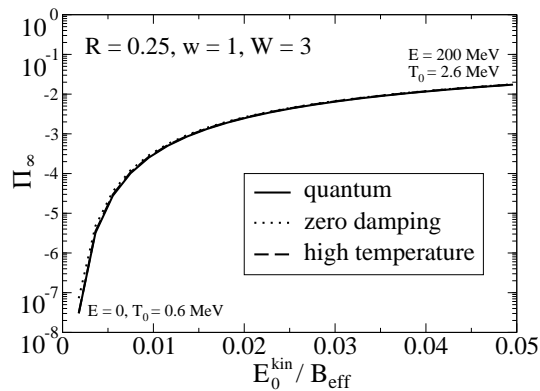}
\caption{\label{fig-limPiofE-macro} Same quantities as in the right
  panel of Fig.~\ref{fig-limPiofE-micro} for macroscopic transport
  coefficients. The value $E = 10~\textrm{MeV}$ corresponds to
$E_{0}^\textrm{kin}/B_\textrm{eff} = 0.0036$.  
Note the logarithmic scale.}
\end{center} \end{figure}

In Fig.~\ref{fig-limPiofE-macro} we show $\Pi_\infty$ as a
function of the ratio $E_{0}^\textrm{kin}/B_\textrm{eff}$ on a logarithmic
scale for $E = 0 \ldots 200 ~\textrm{MeV}$ and $R = 0.25$, $w = 1$
and $W = 3$ using macroscopic transport coefficients now.
The back-bending found in Fig.~\ref{fig-limPiofE-micro} is absent
here. Quantum effects of the collective motion qualitatively
behave as discussed in Sect.~\ref{microresults}. Quantitatively at
$T = 0.8 \ldots 1.0 ~\textrm{MeV}$ they are even smaller than in
the case of microscopic transport coefficients, comp.
Fig.~\ref{fig-alpha2}. The small differences between the full
quantum calculation (of the diffusion coefficients) and the limits
of zero damping and high temperature are hardly visible on
logarithmic scale.

Due to the fact that for macroscopic transport coefficients
collective motion is strongly over-damped ($\eta_b = 5.9 \ldots
4.9$ for $T = 0.6 \ldots 3 ~\textrm{MeV}$) the effective barrier
height (\ref{barreff}) is {\em very} large now: $B_\textrm{eff} = 1.4
~\textrm{GeV} \ldots 980 ~\textrm{MeV}$. Mind that for microscopic
transport coefficients the same quantity is in the range
$B_\textrm{eff} = 15 \ldots 180 ~\textrm{MeV}$ only (comp. $E_{b}
= 10 ~\textrm{MeV}$). This leads to ratios
$E_{0}^\textrm{kin}/B_\textrm{eff}$ of about one order of
magnitude smaller than for microscopic transport coefficients. This
ratio crucially influences the size of the formation probability
$\Pi_\infty$. Consequently in the case of macroscopic transport
coefficients $\Pi_\infty$ is several orders of magnitudes smaller than
for microscopic ones. This effect increases with decreasing $E$.
Compared to this loss of orders of magnitude in $\Pi_\infty$ due to the
replacement of microscopic by macroscopic transport coefficients,
quantum effects in the collective motion due to the type of
approximation applied in the calculation of the
$\Sigma_{\mu\nu}^{eq,b}$ are of minor importance only.

\begin{figure}[hbt] \begin{center}
\epsfig{file=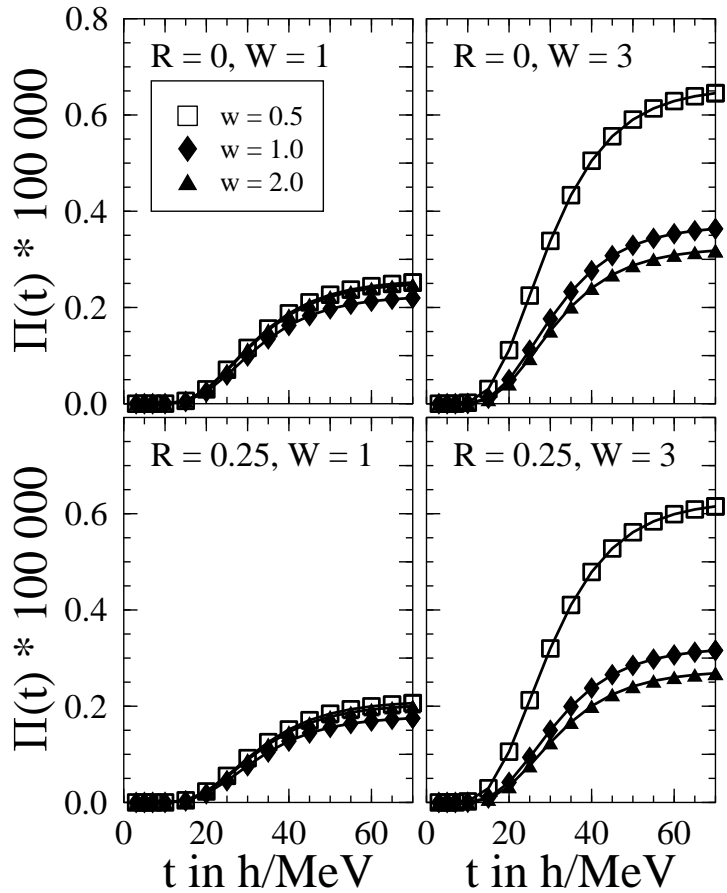, width=0.7\linewidth}
\caption{\label{fig-macro-Pioft} Same quantities as in
Fig.~\ref{fig-micro-Pioft} for $E = 10~\textrm{MeV}$ and macroscopic
transport coefficients.}
\end{center} \end{figure}
The time dependent behavior of the formation probability $\Pi(t)$
is shown in Fig.~\ref{fig-macro-Pioft} for the same initial
conditions as for Fig.~\ref{fig-micro-Pioft}.
As mentioned previously, now the effective damping of the collective
motion is clearly in the over-damped regime: $\eta_{b} \approx
5.8$ both at $T = 0.94~\textrm{MeV}$ ($R = 0$) and 
$T = 0.82~\textrm{MeV}$ ($R = 0.25$). The time scale for the rise of
$\Pi(t)$ is larger by a factor of the order of $10$, as compared
to the microscopic case and completely independent of the initial
kinetic energy $E_{0}^\textrm{kin}$, viz the initial momentum
$P_{0}$. We have checked that different to the case of microscopic
transport coefficients this is still true at $E = 50~\textrm{MeV}$. 
This is in accord with the fact that for $\eta_b \gg 1$ one is in the
Smoluchowski limit, for the momentum drops out entirely. For the
same reason also the stationary value $\Pi_\infty$ is much less
sensitive to the initial kinetic energy than in the case of
microscopic transport coefficients. Whereas in the microscopic
case at $E = 10~\textrm{MeV}$ we gained two orders of magnitude 
going from $R = 0$ to $R = 0.25$ the increase of $\Pi_\infty$ now 
is invisibly small. Using the large energy $E = 50~\textrm{MeV}$ 
the corresponding increase is at most by a factor of $2$ whereas in 
the microscopic case we gain two orders of magnitude, too.

\section{Summary and conclusion}
\label{conclusion}

Concentrating on the formation probability we have been able to
demonstrate the importance of microscopic transport
coefficients of average motion. As compared to those of
macroscopic pictures like that of the liquid drop model and wall
friction, they may increase the
formation probabilities by a few orders of magnitude. This is
especially important at smaller temperatures where the dynamics of
nucleons in a mean field is dominated by quantum effects and where
"collisional damping" is suppressed \cite{hoh:pr:97}. For
collective motion this implies a moderate to weak damping.

In contrast to this, quantum effects in collective motion would
become relevant only at even smaller temperatures. In a sense,
this is in accord with the findings in \cite{hoh.ivf:prl:99} for nuclear
fission. Since at small $T$ friction is suppressed quantal
diffusion coefficients may be estimated within the so called "zero
friction limit". This feature applies only for the microscopic
picture of nucleonic dynamics. In fact, in this paper it was the
first time that microscopic transport coefficients have been used
when calculating the diffusion coefficients for collective
dynamics through the quantal fluctuation dissipation theorem.
Because of the delicate temperature dependence of the transport
coefficients this procedure is vital to get decent answers.

Another issue of our studies was to examine the relevance of the
initial conditions for the phase space distribution. Whenever the
whole process of fusion is split into parts where the entrance
phase is treated differently from the formation of the compound
nucleus uncertainties for the initial conditions of the formation
process are more or less inherent. Evidently, such a situation would
only be given if the intermediate step would correspond to a kind of
quasi-equilibrium; in remote sence in analogy to N. Bohr's hypothesis
of a two step process. 
Indeed, as we have seen, for a more quantitative calculation of the 
reaction cross section it will be inevitable to follow the whole process,
entrance phase plus formation. It is only then that the "initial
conditions" for the latter can be known better. Unfortunately,
however, it is only fair to say that a decent microscopic picture
for the first stage is still lacking.

\vspace{5mm}
\paragraph{Acknowledgments:}
The authors would like to thank F.A.~Ivanyuk for critical and fruitful
discussions and D.~Boilley for helpful comments.

\begin{appendix}

\section{Average coordinate and fluctuations}
\label{timeevol}

Using dimensionless coordinates and the conjugate momenta 
(\ref{dimless-QP}) the solution of the equations of motion for the 
average trajectory reads in the case of the inverted oscillator 
\cite{hoh:pr:97}:
\bal{sol-q}
q(t) & = & \frac{1}{2{\cal E}_{b}} \,( \varpi_{b} P_{0}/\hbar - q_{0} z^{-})
\,\exp(z^{+} (t - t_{0})) \nonumber \\
& + & \frac{1}{2{\cal E}_{b}} \,(-\varpi_{b} P_{0}/\hbar + q_{0} z^{+})
\,\exp(z^{-} (t - t_{0}))
\end{eqnarray}
Here $q_{0} = q(t_{0})$ and $P_{0} = P(t_{0})$ are the initial
coordinate and momentum respectively and the frequencies $z^{\pm}$
are defined in (\ref{zpm}); ${\cal E}_{b} = (2\tau_\textrm{kin})^{-1}
\sqrt{1 + \eta_b^{-2}}$. For long times $t \gg 1/|z^{-}|$
(\ref{sol-q}) reduces to
\bel{lim-sol-q}
q(t) \sim \frac{1}{2{\cal E}_{b}} \,( \varpi_{b} P_{0}/\hbar - q_{0} z^{-})
\,\exp(z^{+} (t - t_{0})) \,.
\end{equation}

In dimensionless coordinates the fluctuations of the coordinate of the
inverted oscillator read (comp. \cite{hoh:pr:97}):
\bal{sol-fluct}
\Sigma_{qq}(t) & = &
A_{1} \,\frac{\varpi_{b}^{2}}{\hbar z^{-}} \,\exp(2z^{-}(t - t_{0})) +
A_{2} \,\frac{\varpi_{b}^{2}}{\hbar z^{+}} \,\exp(2z^{+}(t - t_{0}))
\nonumber \\
& & + A_{3} \,\frac{\varpi_{b}}{\hbar \eta_b}
\,\exp((z^{-} + z^{+}) (t - t_{0})) + \Sigma_{qq}^{eq,b}
\end{eqnarray}
Here the abbreviations
\begin{eqnarray}
A_{1} & = & \frac{z^{-}}{(z^{+} - z^{-})^{2}}
\,\left[ -2 \,\frac{z^{+}}{\varpi_{b}} \,c
+ \left( \frac{z^{+}}{\varpi_{b}} \right)^{2} \,b + a \right]
\label{def-A1} \\
A_{2} & = & \frac{z^{+}}{(z^{+} - z^{-})^{2}}
\,\left[ -2 \,\frac{z^{-}}{\varpi_{b}} \,c
+ \left( \frac{z^{-}}{\varpi_{b}} \right)^{2} \,b + a \right]
\label{def-A2} \\
A_{3} & = & \frac{z^{+} + z^{-}}{(z^{+} - z^{-})^{2}}
\,\left[ -\frac{z^{-} + z^{+}}{\varpi_{b}} \,c
+ \frac{z^{-} z^{+}}{\varpi_{b}^{2}} \,b + a \right]
\label{def-A3}
\end{eqnarray}
and
\bel{def-abc}
\left( \begin{array}{c} a \\ b \\ c \end{array} \right) =
\left( \begin{array}{c} (\Sigma_{pp}(t_{0}) - \Sigma_{pp}^{eq,b}) / \hbar \\
(\Sigma_{qq}(t_{0}) - \Sigma_{qq}^{eq,b}) \hbar \\ \Sigma_{qp}(t_{0})
\end{array} \right)
\end{equation}
have been applied (note the small differences in the definitions
compared to \cite{hoh:pr:97} that make the expressions more symmetric).
The long time limit of (\ref{sol-fluct}) reads:
\bel{lim-sol-fluct} \Sigma_{qq}(t) \sim A_{2}
\,\frac{\varpi_{b}^{2}}{\hbar z^{+}} \,\exp(2z^{+}(t - t_{0}))
\end{equation}

\section[Fluctuation dissipation theorem]
{Implications of the fluctuation dissipation theorem}
\label{diffcoeff-FDT}

For a stable system the equilibrium fluctuations of some quantity 
$A_{\mu}$ can be calculated from the quantal fluctuation dissipation 
theorem in its normal form: 
\bel{FDT}
\Sigma_{\mu\nu} = \hbar \int \frac{d\omega}{2\pi} 
\,\coth \frac{\hbar\omega}{2T} \,\chi_{\mu\nu}''(\omega)
\end{equation}
Here, $\chi_{\mu\nu}''$ is the dissipative part of the
response function that describes the variation of the average 
of $A_{\mu}$ to a small change of the ``external field'' $a_{\nu}$ 
which couples to the system through an interaction 
$\delta H = \sum_{\nu} A_{\nu} a_{\nu}(t)$. 
For the present case, the average $\langle A_{\mu} \rangle$ either
refers to the coordinate $q$ or the momentum $p$.

Unstable modes, which appear in the barrier region, may be treated by
analytic continuation \cite{hoh.kid:ijmpe:98}, changing a positive
stiffness into a negative one. In this way the stationary fluctuations 
of the damped inverted oscillator (of frequency $\varpi_{b}$ and 
effective damping $\eta_{b}$) become:
\begin{eqnarray}
\Sigma_{qq}^{eq,b} & = & -\frac{T}{\hbar\varpi_{b}} 
\left( 1 - \frac{\hbar\varpi_{b}^{2}}{\pi T} \times
\right. \nonumber \\
& & \left. \sum_{j=1}^{3} \frac{\varpi_{D} - \imag\omega^{(j)}}
{(\omega^{(j)} - \omega^{(j+1)}) (\omega^{(j)} - \omega^{(j+2)})} 
\Psi \left( 1 + \imag \,\frac{\hbar\omega^{(j)}}{2\pi T} \right) \right)
\label{Sqqeq} \\
\Sigma_{pp}^{eq,b} & = & \hbar^{2} \left( |\Sigma_{qq}^{eq,b}| 
+ \frac{\varpi_{D}}{\pi\tau_\textrm{kin}\varpi_{b}} \times
\right. \nonumber \\
& & \left. \sum_{j=1}^{3} \frac{-\imag\omega^{(j)}}
{(\omega^{(j)} - \omega^{(j+1)}) (\omega^{(j)} - \omega^{(j+2)})} 
\Psi \left( 1 + \imag \,\frac{\hbar\omega^{(j)}}{2\pi T} \right) \right)
\label{Sppeq} \\
\Sigma_{qp}^{eq,b} & = & 0
\label{Sqpeq}
\end{eqnarray}
Here $\varpi_{D} \gg \varpi_{b}$ is the so called Drude cut-off
frequency necessary to regularize the $\omega$-integral in (\ref{FDT})
in the case of $\Sigma_{pp}^{eq,b}$. We have put 
$\varpi_{D} = 10~\textrm{MeV}/\hbar$ in this paper 
and the frequencies $\omega^{(j)}$ are the solutions of the equation
\bel{eqn-omegaj}
- \varpi_{b}^{2} 
- \imag\omega \left( -\frac{\varpi_{b}^{2}}{\varpi_{D}} 
+ \frac{1}{\tau_\textrm{kin}} \right) - \omega^{2} 
+ \imag \,\frac{\omega^{3}}{\varpi_{D}} = 0
\end{equation}
(periodicity is assumed: $\omega^{(j+3)} \equiv \omega^{(j)}$).

\end{appendix}


\begin{thebibliography}{10}

\bibitem{ann.che.naa.pev.vov:prc:95}
N. Antonenko, E. Cherepanov, A. Nasirov, V. Permjakov, and V. Volkov,
  Phys.~Rev. {\bf C 51},  2635  (1995);
G. Adamian, N. Antonenko, W. Scheid, and V. Volkov, Nucl.~Phys. {\bf A 633},
  409  (1998).

\bibitem{dev.hos:prc:00}
V.~Y. Denisov and S. Hofmann, Phys.~Rev. {\bf C 61},  034606  (2000).

\bibitem{aby:epja:02}
Y. Abe, Eur. Phys. J. {\bf A 13},  143  (2002).

\bibitem{shc.kog.aby:prc:02}
C. Shen, G. Kosenko, and Y. Abe, Phys.~Rev. {\bf C 66},  061602  (2002).

\bibitem{kog.shc.aby:jnrs:02}
G. Kosenko, C. Shen and Y. Abe, J.~Nucl.~Radiochem.~Sci. {\bf 3}, 19 (2002).

\bibitem{ary.ohm:pan:03}
Y. Aritomo and M. Ohta, Phys.~Atom.~Nucl. {\bf 6},  1  (2003).

\bibitem{hoh.ivf:prl:99}
H. Hofmann and F.~A. Ivanyuk, Phys. Rev. Lett. {\bf 82},  4603  (1999).

\bibitem{hoh.ivf.ruc.yas:prc:01}
H. Hofmann, F.~A. Ivanyuk, C. Rummel, and S. Yamaji, Phys.~Rev. {\bf C 64},
  054316  (2001).

\bibitem{hoh.sip:plb:75}
H. Hofmann and P. Siemens, Phys.~Lett. {\bf B 58},  417  (1975).

\bibitem{hoh:pr:97}
H. Hofmann, Phys. Rep. {\bf 284 (4\&5)},  137  (1997).

\bibitem{hoh.kid:ijmpe:98}
H. Hofmann and D. Kiderlen, Int. J. Mod. Phys. {\bf E 7},  243  (1998).

\bibitem{ivf.hoh.pav.yas:prc:97}
F.~A. Ivanyuk, H. Hofmann, V. Pashkevich, and S. Yamaji, Phys.~Rev. {\bf C 55},
   1730  (1997).

\bibitem{pok.hoh:jp:81}
K. Pomorski and H. Hofmann, J. Physique {\bf 42},  381 (1981).

\bibitem{aby.bod.gib.wat:pre:00}
Y. Abe, D. Boilley, B.~G. Giraud, and T. Wada, Phys. Rev. {\bf E 61},  1125
  (2000).

\bibitem{ary.wat.ohm.aby:prc:99}
Y. Aritomo, T. Wada, M. Ohta, and Y. Abe, Phys. Rev. {\bf C 59},  796  (1999).

\end{thebibliography}
\end{document}